\documentclass[runningheads]{svmult}
\usepackage{graphicx}  
\usepackage{physprbb}  
\usepackage{sidecap}   

\newcommand{\bea}{\begin{eqnarray}}
\newcommand{\eea}{\end{eqnarray}}
\newcommand{\nn}{\nonumber}
\newcommand{\nnl}{\nonumber\\}

\renewcommand{\appendix}{
  \setcounter{section}{0}
  \renewcommand{\thesection}{\Alph{section}}
}
\begin{document}
\title*{Correlations and Equilibration in Relativistic
Quantum Systems\footnote{supported by GSI Darmstadt}}
\toctitle{Correlations and Equilibration in Relativistic
Quantum Systems}
%
%
\titlerunning{Correlations and Equilibration}
%
\author{W. Cassing \and S. Juchem}
\authorrunning{W. Cassing and S. Juchem}
%
%
\institute{Institut f\"ur Theoretische Physik,
 Universit\"at
Giessen, Heinrich-Buff-Ring 16, D-35392 Giessen, Germany}

\maketitle              
\begin{abstract}
In this article we study the time evolution of an interacting
field theoretical system, i.e. $\phi^4$-field theory in 2+1
space-time dimensions, on the basis of the Kadanoff-Baym
equations for a spatially homogeneous system including the
self-consistent tadpole and sunset self-energies. We find that
equilibration is achieved only by inclusion of the sunset 
self-energy. Simultaneously, the time evolution of the scalar particle
spectral function is studied for various initial states. We also
compare associated solutions of the corresponding Boltzmann
equation to the full Kadanoff-Baym theory. This comparison shows
that a consistent inclusion of the spectral function has a
significant
impact on the equilibration rates only if the width of the
spectral function becomes larger than 1/3 of the particle mass.
Furthermore, based on these findings, the conventional transport
of particles in the on-shell quasiparticle limit is extended to
particles of finite life time by means of a dynamical spectral function
$A(X,\vec{p},M^2)$.  The off-shell propagation is implemented
in the Hadron-String-Dynamics (HSD) transport code and applied 
to the dynamics of nucleus-nucleus collisions.
\end{abstract}
%
%
%
%

\section{Introduction}
The many-body theory of strongly interacting particles out of
equilibrium is a challenging problem since a couple of decades.
Many approaches based on the Martin-Schwinger hierarchy of Green's
functions \cite{js61} have been formulated
\cite{Wang1,CaWa,cs85,md90,mh94} and applied to model cases.
Nowadays, the dynamical description of strongly interacting
systems out of equilibrium is dominantly based on on-shell transport
theories and efficient numerical recipies have been set up for the
solution of the coupled channel transport equations
\cite{Stoecker,Bertsch,CMMN,Cass,Fuchs,URQMD,CB99} (and Refs.
therein). These transport approaches have been derived either from
the Kadanoff-Baym equations \cite{kb62} in Refs.
\cite{pd841,Bot,Mal,ph95,gl98} or from the hierarchy of connected
equal-time Green's functions \cite{Wang1,Zuo} in Refs.
\cite{CaWa,Cass,CNW} by applying a Wigner transformation and
restricting to first order in the derivatives of the phase-space
variables ($X, p$).

However, as recognized early in these derivations \cite{CaWa,Mal},
the on-shell quasiparticle limit, that invoked additionally a
reduction of the $8N$-dimensional phase-space to $7N$ independent
degrees of freedom, where $N$ denotes the number of particles in
the system, should not be adequate for particles of short life
time and/or high collision rates. Therefore, transport
formulations for quasiparticles with dynamical spectral functions
have been presented in the past \cite{ph95} providing a
formal basis for an extension of the presently applied transport
models. Apart from the transport
extensions mentioned above there is a further branch discussing
the effects from non-local collisions terms (cf. Refs.
\cite{Rudy1,Rudy2,Pavel,Mora1,Mora2,Mora3}).

The basic questions in this field are: \\ i) What are the microscopic
mechanisms that lead to thermalization though the underlying
quantum theory is time reversible? \\ ii) What are the effects of
off-shell transitions in case of strong coupling? \\
iii) Is there an
adequate semiclassical limit to the full quantum theory?

In this article we will attempt to shed some light on these
questions for fully relativistic self-interacting systems and
review some recent progress achieved in the last years. Our work
is organized as follows: In Section 2 we will study numerically
the self-interacting scalar $\phi^4$-theory in 2+1 dimensions on
the basis of the Kadanoff-Baym equations for homogeneous systems
and concentrate on the time evolution of correlations and the
single-particle spectral function. A comparison to the respective
Boltzmann limit is performed in Section 3. Furthermore, the
derivation of a semiclassical off-shell transport theory
for inhomogeneous systems is presented in Section 4 with special
emphasis on the time evolution of spectral functions.

\section{The Kadanoff-Baym Equations}
We briefly recall the
basic equations for Green's functions and particle self-energies as
well as their symmetry properties that will be exploited throughout this work.

\subsection{Preliminaries}
Within the framework of the {\sl closed-time-path}
formalism \cite{js61,kb62} Green's functions $G$ and
self-energies $\Sigma$ are given as path ordered quantities. They
are defined on the time contour consisting of two branches from
(+) $- \infty$ to $\infty$ and (--) from $\infty$ to $- \infty$.
For convenience these propagators and self-energies are
transformed into a $2\times 2$ matrix representation according to
their path structure \cite{cs85}, i.e. according to the
chronological $(+)$ or antichronological $(-)$ branch for the
time coordinates $x_0$ and $y_0$.
Explicitly the Green's functions are given by
\bea
i \, G^{c}_{xy} & = & i \, G^{++}_{xy} \: = \:
\langle \, T^c \, \{ \, \phi(x) \: \phi^{\dagger}(y) \, \} \, \rangle \, , \nnl
i \, G^{<}_{xy} & = & i \, G^{+-}_{xy} \: = \:
\; \eta \: \; \langle \, \{ \, \phi^{\dagger}(y) \: \phi(x) \, \} \, \rangle \, , \nnl
i \, G^{>}_{xy} & = & i \, G^{-+}_{xy} \: = \:
\; \phantom{\eta} \: \; \langle \, \{ \, \phi(x) \: \phi^{\dagger}(y) \, \} \, \rangle \, ,\nnl
i \, G^{a}_{xy} & = & i \, G^{--}_{xy} \: = \:
\langle \, T^a \, \{ \, \phi(x) \: \phi^{\dagger}(y) \, \} \, \rangle .
\label{green_def}
\eea
where the subscript $\cdot_{xy}$ denotes the dependence on the
coordinate space variables $x$ and $y$ and $T^c \, (T^a)$
represent the (anti-)time-ordering operators. In the definition of
$G^<$ the factor $\eta = +1$ stands for bosons and $\eta = -1$ for
fermions. In the following we will consider a theory for scalar
bosons.
The full Green's functions are determined via the
Dyson-Schwinger equations for path-ordered quantities, here
given in $2 \times 2$ matrix representation as
\bea
\left(
\begin{array}{cc}
G^{c}_{\phantom{\!o}} & G^{<}_{\phantom{\!o}} \\[0.2cm]
G^{>}_{\phantom{\!o}} & G^{a}_{\phantom{\!o}}
\end{array}
\right)_{\!\!xy}
\!\!\! =  \,
\left(
\begin{array}{cc}
G^{c}_{\!o} & G^{<}_{\!o} \\[0.2cm]
G^{>}_{\!o} & G^{a}_{\!o}
\end{array}
\right)_{\!\!xy}
\!\!\! + \,
\left(
\begin{array}{cc}
G^{c}_{\!o} & G^{<}_{\!o} \\[0.2cm]
G^{>}_{\!o} & G^{a}_{\!o}
\end{array}
\right)_{\!\!x z}
\!\!\!\!\odot
\left(
\begin{array}{cc}
\Sigma^{c} & -\!\Sigma^{<} \\[0.2cm]
-\!\Sigma^{>} & \Sigma^{a}
\end{array}
\right)_{\!\!z z^{\prime}}
\!\!\!\!\!\odot
\left(
\begin{array}{cc}
G^{c}_{\phantom{\!o}} & G^{<}_{\phantom{\!o}} \\[0.2cm]
G^{>}_{\phantom{\!o}} & G^{a}_{\phantom{\!o}}
\end{array}
\right)_{\!\!z^{\prime}\!y \: .} \!\!\!\!\!
\label{ds_matrix}
\eea
The self-energies $\Sigma^{\cdot}$ are also defined according to
their time structure while the symbol '$\odot$' implies an
integration over the intermediate space-time coordinates from
$-\infty$ to $\infty$. Linear combinations of diagonal and
off-diagonal matrix elements give the retarded and advanced Green's
functions $G^{\rm ret/adv}$ and self-energies $\Sigma^{\rm ret/adv}$
\bea
\begin{array}{ll}
G^{\rm ret}_{xy} \: = \: G^{c}_{xy} \!- G^{<}_{xy}
             \: = \: G^{>}_{xy} \!- G^{a}_{xy} \, , & \qquad
G^{\rm adv}_{xy} \: = \: G^{c}_{xy} \!- G^{>}_{xy}
             \: = \: G^{<}_{xy} \!- G^{a}_{xy} \, , \\[0.5cm]
\Sigma^{\rm ret}_{xy} \: = \: \Sigma^{c}_{xy} \!- \Sigma^{<}_{xy}
                  \: = \: \Sigma^{>}_{xy} \!- \Sigma^{a}_{xy} \, , & \qquad
\Sigma^{\rm adv}_{xy} \: = \: \Sigma^{c}_{xy} \!- \Sigma^{>}_{xy}
                  \: = \: \Sigma^{<}_{xy} \!- \Sigma^{a}_{xy} \, .
\end{array}
\label{retquant}
\eea
Resorting equations (\ref{ds_matrix}) one obtains Dyson-Schwinger
equations for the retarded (advanced) Green's functions (where only
the respective self-energies are involved)
\bea
\hat{G}_{\!o,x}^{-1} \; \; G_{xy}^{\rm ret/adv}
\; \: = \; \: \delta_{xy}
\; \: + \; \: \Sigma_{xz}^{\rm ret/adv} \: \odot \: G_{zy}^{\rm ret/adv} \; ,
\eea
and the wellknown Kadanoff-Baym equation for the Wightman function
$G^<$,
\bea
\hat{G}_{\!o,x}^{-1} \; \; G_{xy}^{<}
\; \: = \; \: \Sigma_{xz}^{\rm ret} \: \odot \: G_{zy}^{<}
\; \: + \; \: \Sigma_{xz}^{<}   \: \odot \: G_{zy}^{\rm adv} \: .
\label{kb_spatial}
\eea
In these equations $\hat{G}_{\!o,x}^{-1}$ denotes the (negative)
Klein-Gordon differential operator which for bosonic field quanta
of mass $m$ is given by
$\hat{G}_{\!o,x}^{-1} = - (\partial^\mu_x \partial^x_\mu + m^{2})$.
The Klein-Gordon equation is solved by the free propagators $G_{\!o}$ as
\bea
\hat{G}_{\!o,x}^{-1} \;
\left( \begin{array}{cc}
G_{\!o}^c & G_{\!o}^< \\[0.2cm]
G_{\!o}^> & G_{\!o}^a
\end{array} \right)_{\!\!xy} \!\!\!\!\!
\; = \; \delta_{xy}
\left( \begin{array}{rr}
1 & 0 \\[0.2cm] 0 & \,-\!1
\end{array} \right) ,
\qquad \quad
\hat{G}_{\!o,x}^{-1} \; \; G_{\!o,xy}^{\rm ret/adv}
\; = \; \delta_{xy} \;
\label{spatial_freeprop}
\eea
with the $\delta$-distribution $\delta_{xy} \equiv \delta^{(d+1)}(x-y)$
for $d+1$ space-time dimensions.

\subsection{Equilibration Within the Scalar $\phi^4$-Theory}
The scalar $\phi^4$-theory is an example for a fully relativistic
field theory of interacting scalar particles that allows to test
theoretical approximations without coming to the problems
of gauge invariant truncation schemes
\cite{Peter,berges1,berges2,berges3}.
Its Lagrangian density is given by $(x=(t,\vec{x}))$
\bea
\label{lagrangian}
{\cal L}(x) \; = \;
  \frac{1}{2} \, \partial_{\mu} \phi(x) \, \partial^{\mu} \phi(x)
\: - \: \frac{1}{2} \, m^2 \, \phi^2(x)
\: - \: \frac{\lambda}{4 !} \, \phi^4(x) \; ,
\eea
where $m$ denotes the mass and $\lambda$ is the coupling
strength determining the interaction strength of the scalar
fields.
Its quantization is performed in the canonical form
(in $d$ spatial dimensions)
\bea
[ \, \phi(x) \, , \, \partial_{y_0} \phi(y) \, ]_{x_0 = y_0}
\; = \; i \, \delta^{(d)}(\vec{x} - \vec{y}) \: .
\label{commutator}
\eea
Since for the real boson theory (\ref{lagrangian}) the relation
$G^{>}(x,y) = G^{<}(y,x)$ holds (\ref{green_def})
the knowledge of the Green's functions $G^{<}(x,y)$
for all $x, y$ characterizes the system completely.
Nevertheless, we will give the equations  for
$G^{<}$ and $G^{>}$ explicitly since this is the familiar
representation for general field theories \cite{md90}.
Self-consistent equations of motion are obtained by a loop
expansion of the two-particle irreducible (2PI) effective action
$\Gamma$ on the closed time path. 
It is given by the Legendre transform of the generating functional 
of the connected Green's functions $W$ as
\bea
\Gamma[G] \; = \;
\Gamma^o \: + \:
\frac{i}{2} \:
\left[ \;
ln ( 1 - \odot_p \, G_o \odot_p \Sigma ) \: + \:
\odot_p \, G \odot_p \Sigma
\;\right] \; + \;
\Phi[G] \: .
\label{effaction}
\eea
in case of vanishing vacuum expectation
value  $\langle 0|\phi(x)|0 \rangle = 0$ \cite{knoll}.
In (\ref{effaction}) $\Gamma^o$ depends only on free Green's
functions and is treated as a constant
while the symbols $\odot_p$ represent convolution
integrals over the closed time path.
The functional $\Phi$ is the sum of all closed $2PI$
diagrams built up by full propagators $G$; it determines
the self-energies by functional variation as
\bea
\Sigma(x,y) \; = \; 2 i \, \frac{\delta \Phi}{\delta G(y,x)} \: .
\eea
Taking into account contributions up to the
3-loop order the $\Phi$ functional reads explicitly for the problem
(\ref{lagrangian})
\bea
i \Phi \; = \; \frac{i \lambda}{8}
\int_{\cal C} d^{d+1\!}x \; \: G(x,x)^2
\; - \;
\frac{\lambda^2}{48}
\int_{\cal C} d^{d+1\!}x
\int_{\cal C} d^{d+1\!}y \; \: G(x,y)^4 .
\label{phifunctional}
\eea
By using the stationarity condition of the action
$ \delta \Gamma / \delta G = 0$ and resolving the
time structure of the path ordered quantities
we obtain the Kadanoff-Baym equations for the time evolution
of the Wightman function \cite{md90,berges1}:
\hspace{-0.7cm}
\bea
\label{kabaeqcs}
\!\!\!\!\!\!\!\!
- \left[
\partial_{\mu}^{x} \partial_{x}^{\mu} \!+ m^2
\right] \, G^{<}(x,y)
& = &
\Sigma_{\rm tad}(x) \; G^{<}(x,y) \nnl[0.3cm]
& + & \!
\int_{t_o}^{x_o} \!\!\!\!\!\!\! dz_o \!\int \!\!d^{d}\!z \;
\left[ \Sigma^{>}(x,z) - \Sigma^{<}(x,z) \right] \: G^{<}(z,y)
\\[0.2cm]
& - & \!
\int_{t_o}^{y_o} \!\!\!\!\!\!\! dz_o \!\int \!\!d^{d}\!z \;\,
\Sigma^{<}(x,z) \: \left[ G^{>}(z,y) - G^{<}(z,y) \right] \!,
\nnl[0.3cm]
\!\!\!\!\!\!\!\!
- \left[
\partial_{\mu}^{y} \partial_{y}^{\mu} \!+ m^2
\right] \, G^{<}(x,y)
& = &
\Sigma_{\rm tad}(y) \; G^{<}(x,y) \nnl[0.3cm]
& + & \!
\int_{t_o}^{x_o} \!\!\!\!\!\!\! dz_o \!\int \!\!d^{d}\!z \;
\left[ G^{>}(x,z) - G^{<}(x,z) \right] \: \Sigma^{<}(z,y)
\nnl[0.2cm]
& - & \!
\int_{t_o}^{y_o} \!\!\!\!\!\!\! dz_o \!\int \!\!d^{d}\!z \;\,
G^{<}(x,z) \: \left[ \Sigma^{>}(z,y) - \Sigma^{<}(z,y) \right]
\!,
\nn
\eea
where $d$ denotes the spatial dimension of the problem ($d=2$ in
the case considered below).
\begin{figure}[t]
\begin{center}
\includegraphics[width=0.7\textwidth]{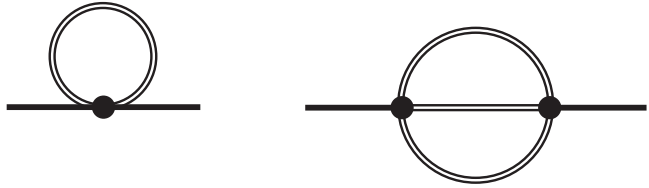}
\end{center}
\caption[]{\it Self-energies of the Kadanoff-Baym equation: tadpole
self-energy (l.h.s.) and sunset self-energy (r.h.s.). Since the
lines represent full Green's functions the self-energies are 
self-consistent (see text)
\label{plot_selfenergy}}
\end{figure}
Within the 3-loop approximation for the effective action
 (i.e. the $\Phi$ functional (\ref{phifunctional}))
we get two different self-energies:
In leading order of the coupling constant only the tadpole diagram 
(l.h.s. of Fig. \ref{plot_selfenergy}) contributes and leads to the
generation of an effective mass for the field quanta. This 
self-energy (in coordinate space) is given by
\bea
\label{tadpole_cs}
\Sigma_{\rm tad}(x) \; = \; \frac{\lambda}{2} \; i \: G^{<}(x,x) \; ,
\eea
and is local in space and time.
In next order in the coupling constant
(i.e. $\lambda^2$)
the non-local sunset self-energy (r.h.s. of Fig. \ref{plot_selfenergy})
enters the time evolution as
\bea
\label{sunset_cs}
\Sigma^{</>}(x,y) \; = \; - \frac{\lambda^2}{6}
\; G^{</>}(x,y) \; G^{</>}(x,y) \; G^{>/<}(y,x) \\[0.3cm]
\longrightarrow \quad
\Sigma^{</>}(x,y) \; = \; - \frac{\lambda^2}{6} \;
\left[ \, G^{</>}(x,y) \, \right]^3 \, .
\eea
Thus the Kadanoff-Baym equation (\ref{kabaeqms}) in our
case includes the influence of a mean-field on the particle
propagation -- generated by the tadpole diagram -- as well
as of scattering processes as inherent in the sunset diagram.

The Kadanoff-Baym equation describes the full quantum
nonequilibrium time evolution on the two-point level for a system
prepared at an initial time $t_0$, i.e. when higher order
correlations are discarded.  The causal structure of this initial
value problem is obvious since the time integrations are performed
over the past up to the actual time $x_0$ (or $y_0$, respectively)
and do not extend to the future.

In the following we will restrict to homogeneous systems in space.
To obtain a numerical solution of the Kadanoff-Baym
equation (\ref{kabaeqcs}) is transformed to momentum space:
\bea
\label{kabaeqms}
\partial^2_{t_1} \, G^{<}(\vec{p},t_1,t_2)
& = &
- [ \, \vec{p}^{\,2} + m^2 + \bar{\Sigma}_{\rm tad}(t_1) \, ]
\; G^{<}(\vec{p},t_1,t_2) \\[0.3cm]
& - &
\int_{t_0}^{t_1} \!\!\! dt^{\prime} \;
\left[ \,
\Sigma^{>}(\vec{p},t_1,t^{\prime}) - \Sigma^{<}(\vec{p},t_1,t^{\prime})
\, \right]
\; G^{<}(\vec{p},t^{\prime},t_2) \nnl[0.1cm]
& + &
\int_{t_0}^{t_2} \!\!\! dt^{\prime} \;
\Sigma^{<}(\vec{p},t_1,t^{\prime}) \;
\left[ \,
G^{>}(\vec{p},t^{\prime},t_2) - G^{<}(\vec{p},t^{\prime},t_2) \,
\right] \nnl[0.2cm]
& = &
- [ \, \vec{p}^{\,2} + m^2 + \bar{\Sigma}_{\rm tad}(t_1) \, ]
\; G^{<}(\vec{p},t_1,t_2) 
\; + \; I_1^{<}(\vec{p},t_1,t_2)
\nn
\eea
where we have summarized both memory integrals into the 
function $I_1^{<}$.
The equation of motion in the second time direction $t_2$
is given analogously.
In two-time and momentum space ($\vec{p},t,t'$)
representation the self-energies read
\bea
\label{sems}
\bar{\Sigma}_{\rm tad}(t)
& = &
\frac{\lambda}{2} \, \int \!\! \frac{d^{d\!}p}{(2\pi)^d} \; \;
i \, G^{<}\!(\vec{p},t,t) \; ,
\\[0.4cm]
\Sigma^{<}\!(\vec{p},t,t^{\prime})
& = &
- \frac{\lambda^2}{6}
\int \!\! \frac{d^{d\!}q}{(2\pi)^{d}} \!
\int \!\! \frac{d^{d\!}r}{(2\pi)^{d}} \;
G^{<}\!(\vec{q} ,t,t^{\prime}) \;
G^{<}\!(\vec{r} ,t,t^{\prime}) \;
G^{>}\!(\vec{q} \!+\! \vec{r} \!-\! \vec{p} ,t^{\prime}\!,t) \, . \nnl[0.2cm]
& = &
- \frac{\lambda^2}{6}
\int \!\! \frac{d^{d\!}q}{(2\pi)^{d}} \!
\int \!\! \frac{d^{d\!}r}{(2\pi)^{d}} \;
G^{<}\!(\vec{q} ,t,t^{\prime}) \;
G^{<}\!(\vec{r} ,t,t^{\prime}) \;
G^{<}\!(\vec{p} \!-\! \vec{q} \!-\! \vec{r} ,t,t^{\prime}) \, .
\nn
\eea

\subsection{Renormalization}
In 2+1 space-time dimensions both self-energies incorporated
(\ref{plot_selfenergy}) are ultraviolet divergent.
Since we consider particles with a finite mass no problems
arise from the infrared momentum regime.
For the renormalization of the divergences we only suppose
that the time-dependent nonequilibrium distribution functions
are decreasing for large momenta comparable to the equilibrium ones,
i.e exponentially.
Thus we can apply the usual finite temperature renormalization
scheme.
By separating the real time (equilibrium) Green's functions into
vacuum ($T=0$) and thermal parts it becomes apparent that only
the pure vacuum contributions of the self-energies are divergent.
For the linear divergent tadpole diagram we introduce a mass
counterterm (at renormalized mass $m$) as
\bea
\delta\!m^2_{\rm tad}\; = \; \int \frac{d^2\!p}{(2\pi)^2} \;
\frac{1}{2 \omega_{\vec{p}}} \; ,
\qquad \qquad
\omega_{\vec{p}} = \sqrt{\vec{p}^2 + m^2} \; ,
\eea
that cancels the contribution from the momentum integration
of the vacuum part of the Green's function.
In case of the sunset diagram only the logarithmically divergent
pure vacuum part requires renormalization, while it remains
finite as long as at least one temperature line is involved.
Contrary to the case of 3+1 dimensions it is not necessary to
employ the involved techniques developed for the renormalization of
self-consistent theories (in equilibrium) in
Refs. \cite{knollren}.
Since the divergence only appears (in energy-momentum space) in
the real part of the Feynman self-energy $\Sigma^{c}$ at $T=0$
(and equivalently in the real part of the retarded/advanced 
self-energies $\Sigma^{\rm ret/adv}$), it can be absorbed by 
another mass counterterm
\bea
\delta\!m^2_{\rm sun} & = &
- {\rm Re\,}\Sigma^{c}_{T=0}(p^2)
\: = \: - {\rm Re\,}\Sigma^{\rm ret/adv}_{T=0}(p^2) \\[0.2cm]
& = &
\frac{\lambda^2}{6}
\int \!\!\! \frac{d^2\!q}{(2 \pi)^2} \,
\int \!\!\! \frac{d^2\!r}{(2 \pi)^2} \; \,
\frac{1}{4 \,\omega_{\vec{q}} \, \omega_{\vec{r}} \, \omega_{\vec{q}+\vec{r}-\vec{p}}} \;
\;
\frac{ \omega_{\vec{q}}\!+\!\omega_{\vec{r}}\!+\!\omega_{\vec{q}+\vec{r}-\vec{p}} }
     { [ \, \omega_{\vec{q}}\!+\!\omega_{\vec{r}}\!+\!\omega_{\vec{q}+\vec{r}-\vec{p}} \, ]^2 - p_0^2 }
\nn
\eea
at given 4-momentum $p=(p_0,\vec{p})$ and renormalized mass $m$.

\subsection{Numerical implementation}
For the solution of the Kadanoff-Baym equations we generate a closed set
of first order differential equations in time for the Green's functions
$ i\,G_{\phi \phi}^{<}(x,y)
\,=\,\langle\, \phi(y)\,\phi(x) \,\rangle
$,
$i\,G_{\pi \phi}^{<}(x,y)
\,=\, \langle\, \phi(y)\,\pi(x) \,\rangle$
and
$i\,G_{\pi \pi}^{<}(x,y)
\,=\, \langle\, \pi(y)\,\pi(x) \,\rangle $
with the canonical field momentum $\pi(x) = \partial_{x_0} \phi(x)$.
The disadvantage of having more Green's functions in this scheme
is compensated by its good accuracy. We especially take into account
the propagation along the time diagonal which leads to an improved
numerical precision. The set of differential equations is solved by
means of a Runge-Kutta algorithm of 4th order. For the
calculation of the self-energies we apply a Fourier method
similar to the one used in \cite{pd841,koe1}.
The self-energies (\ref{sems}) are calculated in coordinate space,
where they are products of coordinate-space Green's functions --
that are available by Fourier transformation -- and are finally
transformed to momentum space again.
We note that the momentum space is discretized by a finite number
of momentum states fixed by periodic boundary conditions in a finite
volume $V= a^2$.

\subsection{Equilibration}
As already observed in the 1+1 dimensional case
\cite{berges1,berges2,berges3} the inclusion of the mean field does not
lead to an equilibration of arbitrary initialized momentum
distributions, since it only modifies the propagation of the
particles by the generation of an effective mass. Thermalization
in 2+1 dimensions requires the inclusion of the collisional 
self-energies from the sunset diagram.
\begin{figure}[hbt]
\vspace{-0.6cm}
\begin{center}
\includegraphics[width=0.52\textwidth]{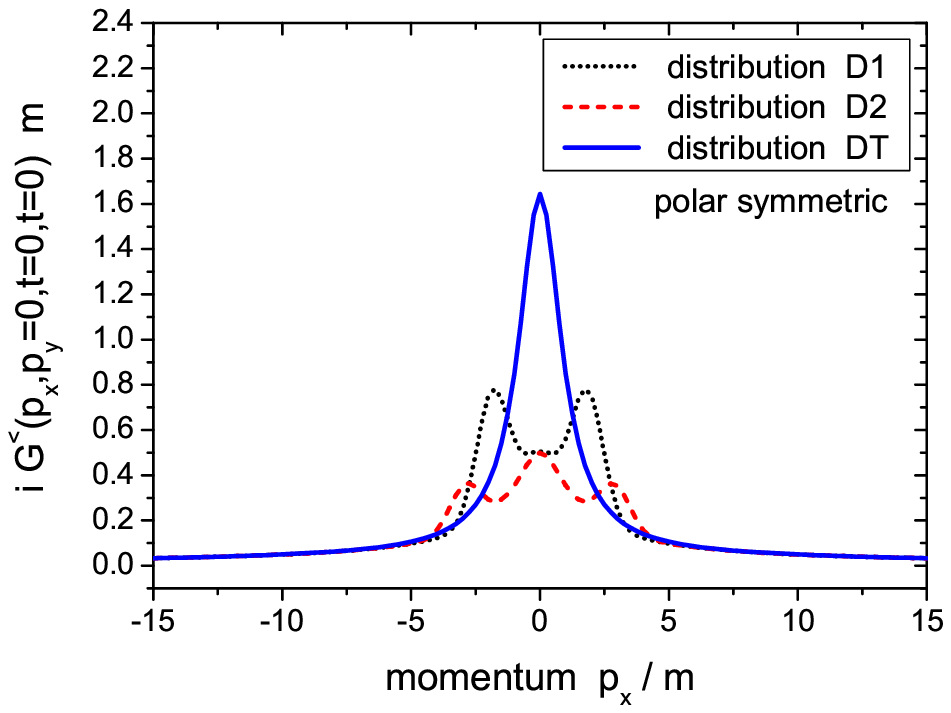}
\hspace{-0.7cm}
\includegraphics[width=0.52\textwidth]{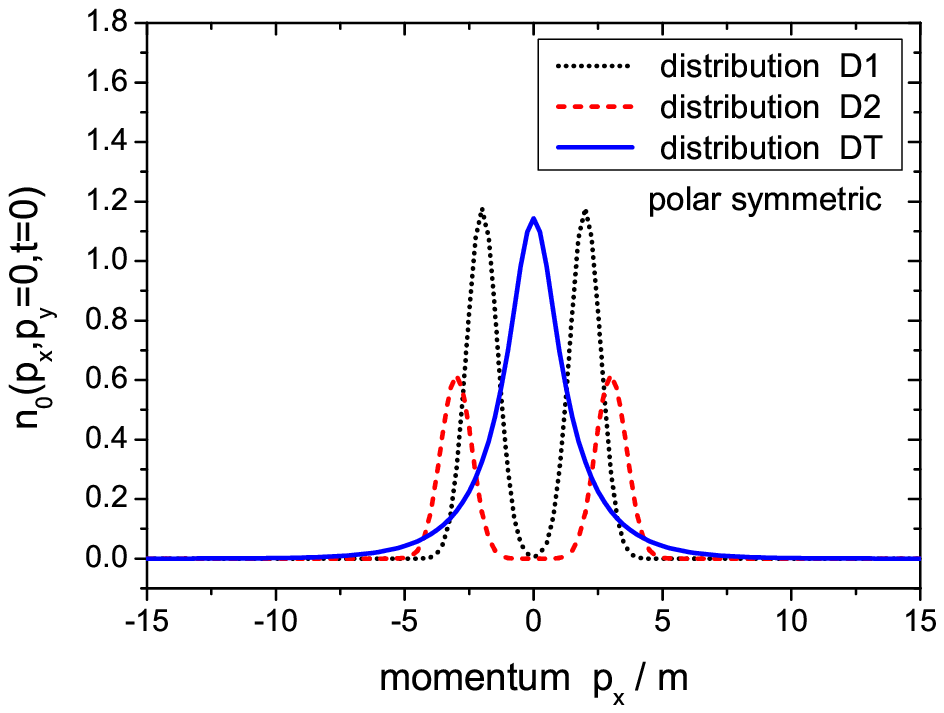}
\end{center}
\vspace{-0.5cm}
\caption[]{\it  Initial Green's functions $
{\it i G^{<}} (|\,\vec{p}\,|,{\it t}\!=\!0,{\it t}\!=\!0)$ 
({\it left}) and corresponding initial
distribution functions ${\it n_o}(|\,\vec{p}\,|,{\it t}\!=\!0)$ 
({\it right}) for the
distributions D1, D2 and DT in momentum space (cut of the polar
symmetric distribution in $p_x$ for $p_y = 0$)
\label{plot_ini01}}
\end{figure}
\begin{figure}[bth]
\vspace{-0.7cm}
\begin{center}
\includegraphics[width=1.0\textwidth]{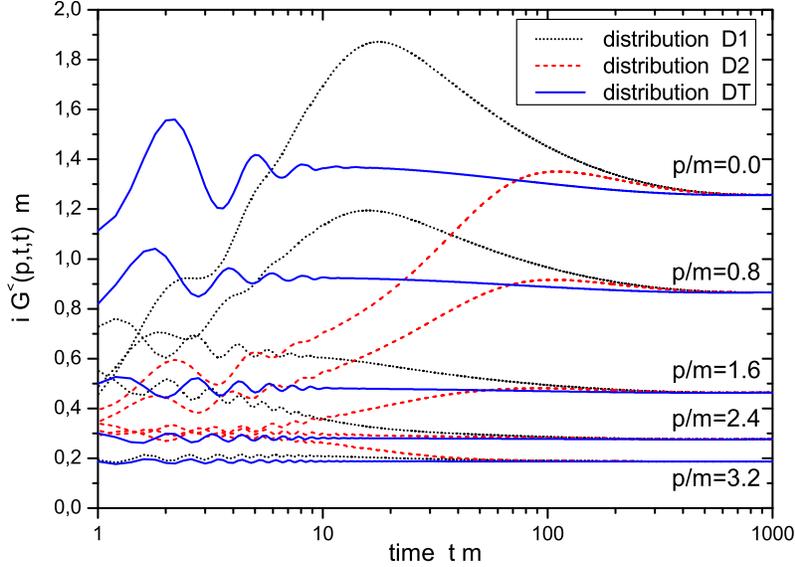}
\end{center}
\vspace{-0.7cm}
\caption[]{\it Time evolution of various momentum modes of the 
equal-time Green's function for 
$|\,\vec{p}\,|/{\it m} = 0.0, 0.8, 1.6, 2.4, 3.2$
(from top to bottom) for three different initial configurations
D1, D2, DT (characterized by the different line type) with the
same energy density. For the rather strong coupling constant
${\it \lambda/m} = 16$ the initial oscillations  are damped 
rapidly and have already disappeared at
${\it t \cdot m} = 20$. Finally all momentum modes assume the same
respective equilibrium value independent of the initial state
\label{plot_equi01}}
\end{figure}
To demonstrate the phenomenon of equilibration we consider the
time evolution of several initial conditions characterized by the
same energy density. The initial equal-time Green's functions $i
G^{<}(\vec{p},t\!=\!0,t\!=\!0)$ adopted are displayed in Fig.
\ref{plot_ini01} (l.h.s.) as a function of the momentum $p_x$ (for
$p_y=0$). We concentrate on polar symmetric configurations due to
the large numerical expense for this first
investigation\footnote{In Section 3 we will present calculations
for non-symmetric systems.}. Since the equal-time Green's functions
$G^{<}(\vec{p},t,t)$ are purely imaginary we display 
(the real part of) $i \, G^{<}$. 
Furthermore, the corresponding initial distribution
functions $n_o(\vec{p},t\!=\!0)$ are shown in Fig. \ref{plot_ini01}
(r.h.s.). While the initial distributions D1, D2 have the shape of
(polar symmetric) 'tsunami' waves with maxima at different
momenta, the initial distribution DT corresponds to a free Bose gas
at a given initial temperature. The difference between the Green's
functions and the distribution functions is given by the vacuum
contribution which has its maximum at small momenta, i.e.
$
2 \omega_{\vec{p}} \, i \, G_{\phi \phi}^{<}(\vec{p},t\!=\!0,t\!=\!0) \; = \;
2 n_o(\vec{p},t\!=\!0) + 1 \ .
$
So even for the distributions D1, D2 the corresponding Green's
functions are nonvanishing at low momentum.
The time evolution of
various momentum modes of the equal-time Green's function for these
three different initial states is shown in Fig. \ref{plot_equi01}.
Starting from different initial conditions (as shown in Fig. 2) the
single momentum modes finally approach the equilibrium values. For
large times all modes reach a static limit as characteristic for
an equilibrated system. On the other hand, for small times one
observes a damped oscillating behaviour. This can be identified as
a typical 'switching-on' effect where the system -- being in a
static (free) situation -- is excited by a sudden increase of the
coupling constant. In the present calculation the initial state is
given by a free state with mass $m = 1$ and the coupling
$\lambda/m = 16$ is switched on for times larger than 0. One
might also start with an effective initial mass due to the 
self-consistent tadpole contribution \cite{berges3}. For the study of
the equilibration process, however, this does not make any
significant difference. Due to the character of the self-energies
the 'switching-on' oscillations are damped in the course of time.
The damping depends on the coupling strength and is much larger
for strongly coupled systems. For very strong interactions the
initial oscillations are even hard to recognize. The final part of
the time evolution is characterized by an approximately
exponential approach to the equilibrium value. In contrast to the
calculations performed for $\phi^4$-theory in 1+1 space-time
dimensions \cite{berges3} there is no intermediate region, where
the momentum modes show a power-law behaviour. Furthermore, we
observe that -- depending on the initial conditions and the coupling
strength -- the momentum modes can temporarily exceed their
respective equilibrium value. This can be seen for the lowest
momentum modes of distribution D1 in Fig. \ref{plot_equi01} which
possesses initially a maximum at relatively small momentum.
Especially the momentum mode $|\vec{p}|=0$ of the equal-time Green's
function $G^{<}$, which starts at around 0.5, is rising to a value
of $\sim$1.8 before it is decreasing again to its equilibrium
value of about 1.26. Thus the evolution towards the final
equilibrium value is -- after the initial phase with damped
oscillations -- not necessarily monotonic. 
%
%
On the other hand, initial
configurations like the distribution DT, where the system
initially is given by a free gas of particles at a temperature
$T_0$, do not show this property. Although they are -- of course
-- not the equilibrium state of the interacting theory, they are
much closer to it than the distributions D1 and D2. Therefore, the
evolution towards the equilibrium distribution runs less
violently.

\subsection{Time Evolution of the Spectral Function}
Within the
Kadanoff-Baym calculations the full quantum character of the
two-point functions is taken into account. Consequently one fully
incorporates the spectral properties of the nonequilibrium system
during its time evolution. The spectral function $A(x,y)$ 
in our case is given by
\bea
A(x,y) \: = \: \langle \, [ \, \phi(x) \, , \, \phi(y) \, ]_{-} \,
\rangle \: = \: i \,
\left[ \, G^{>}(x,y) \: - \: G^{<}(x,y) \, \right] .
\label{spec_def}
\eea
For each system time $T=(t_1\!\!+\!t_2)/2$ the spectral function in
Wigner-space is obtained via Fourier transformation with respect
to the relative time coordinate $\Delta t = t_1\!\!-\!t_2$:
\bea
A(\vec{p}, p_0, T) \: = \:
\int_{-\infty}^{\infty} \!\!\!\! d\Delta t \; \;
e^{i \Delta t \, p_0} \; \;
A(\vec{p}, t_1=T\!+\!\Delta t/2, t_2=T\!-\!\Delta t/2) .
\label{spec_fourier}
\eea
A damping of the function $A(\vec{p},t_1,t_2)$ in relative time
$\Delta t$
corresponds to the generation of a finite width of the spectral
function in Wigner-space. This width in turn can be interpreted as
the inverse life time of the scalar particle. We recall that the spectral
function obeys -- for every mean time $T$ and for all momenta
$\vec{p}$ -- the normalization
\bea
\int_{-\infty}^{\infty} \frac{dp_0}{2 \pi} \;
p_0 \; A(\vec{p},p_0,T) \; = \; 1
\qquad \forall \; \vec{p},\,T
\label{specnorm}
\eea
which is a reformulation of the equal-time commutation relation 
(\ref{commutator}).

\begin{figure}[t]
\begin{center}
\vspace*{-0.3cm}
\hspace*{-0.6cm}
\includegraphics[width=1.1\textwidth]{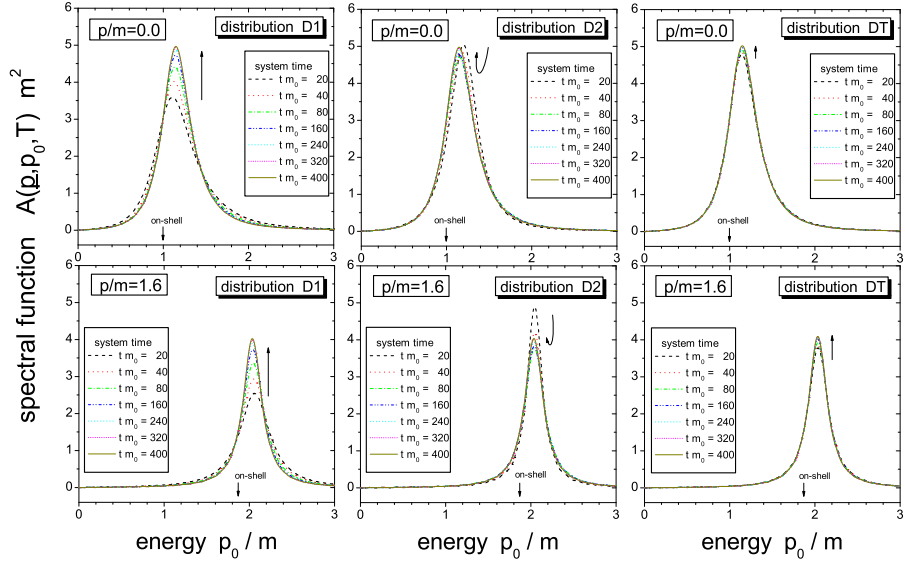}
\end{center}
\vspace{-0.7cm}
\caption[]{\it Time evolution of the spectral function $ A(\vec{p},
p_0, T) $ for the initial distributions D1, D2 and DT (from left
to right) for the two momenta $| \, \vec{p} \, | / m = 0.0$
(upper row) and $| \, \vec{p} \, | / m = 1.6$ (lower row). The
spectral function is shown for several times $t \cdot m =$ 20,
40, 80, 160, 240, 320, 400 as indicated by the different line
types \label{spec01}}
\end{figure}

In Fig. \ref{spec01} we present the time evolution of the spectral
function for the  initial distributions D1, D2 and DT for two
different momentum modes $| \, \vec{p} \, | / m = 0.0$ and $| \,
\vec{p} \, | / m = 1.6$. 
Since the spectral functions are antisymmetric in energy
$A(\vec{p},-p_0,T) = - A(\vec{p},p_0,T)$ we only show the
positive energy part. 
For our initial value problem in two-time space the Fourier 
transformation (\ref{spec_fourier}) is restricted for system 
time $T$ to an interval $\Delta t \in [-2T,2T\,]$.
Thus in the very early phase the spectral function assumes a finite 
width already due to the limited support of the Fourier transformation
and a Wigner representation is not very meaningful.
We therefore present the spectral functions for various system times 
starting from $t \cdot m = 20$ up to $t \cdot m = 400$.

For the free thermal initialization DT the evolution of the
spectral function is very smooth. It is  comparable to the smooth
evolution of the equal-time Green's function for this initialization
as discussed in the last Subsection. The spectral function is already
close to the equilibrium shape at small times being initially only
slightly broader than for late times. The maximum of the
spectral function lies (for all momenta) higher than the (bare)
on-shell value and nearly keeps its position over the whole time
evolution. It results from a positive tadpole mass shift which is
only partly compensated by a downward shift from the sunset
diagram.

The time evolution of the initial distributions D1 and D2 has a
richer structure. For the initial distribution D1 the spectral
function is broad for small system times (see the line for $t
\cdot m = 20$) and is getting a little sharper in the course of
the system evolution (as presented for the momentum mode
$|\,\vec{p}\,| / m = 0.0$ as well as for $|\,\vec{p}\,| / m =
1.6$). At the same time the height of the spectral function is
increasing (as demanded by the normalization property
(\ref{specnorm})) with time. This is indicated by the small arrow
close to the peak position. Furthermore, the maximum of the
spectral function (which is approximately the on-shell energy) is
shifted slightly upwards for the zero mode and downwards for the
mode with higher momentum. Although the real part of the
(retarded) sunset self-energy leads (in general) to a lowering of
the effective mass, the on-shell energy of the momentum modes is
still higher than the one for initial mass $m$
(indicated by the 'on-shell' arrow) due to the positive mass shift
from the tadpole contribution. For the initial distribution D2 we
find an opposite behaviour. Here the spectral function develops at
intermediate times a slightly higher width than in the beginning
before it is approaching the narrower static shape. The
corresponding evolution of the maximum is again indicated by the
(bent) arrow. Finally all spectral functions assume the (same)
equilibrium form.

As already observed above for the equal-time Green's functions, we
emphazise that there is no unique type of evolution for the
nonequilibrium system. In fact, the evolution of the system
during the equilibration process is sensitive to the initial
conditions.

\subsection{Generation of Correlations}

\begin{figure}[t]
\vspace{-0.7cm}
\begin{center}
\includegraphics[width=1.0\textwidth]{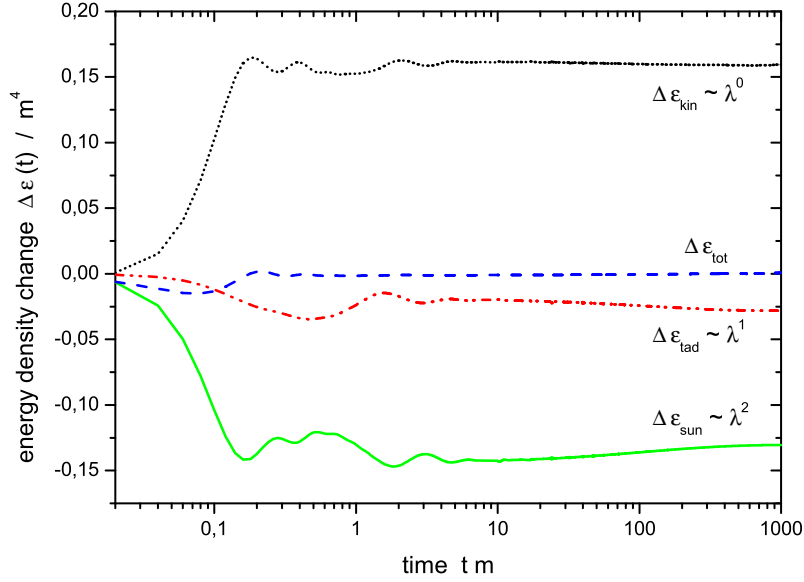}
\end{center}
\vspace{-0.7cm}
\caption[]{\it Change of the different contributions to the total
energy density in time. The sunset energy density
$\varepsilon_{\rm sun}$ decreases as the system correlates. This is
mostly compensated by an increase of the kinetic energy density
$\varepsilon_{\rm kin}$. Together with the smaller tadpole
contribution $\varepsilon_{\rm tad}$ the total energy
$\varepsilon_{\rm tot}$ is conserved
\label{plot_energy01}}
\end{figure}

The time evolution within
the Kadanoff-Baym equation is characterized by the generation of
correlations. This can be seen from Fig. \ref{plot_energy01},
where all energy density contributions are displayed as a function
of time. The kinetic energy density $\varepsilon_{\rm kin}$ is given
by all parts of the total energy density that do not depend on
the coupling constant ($\propto \lambda^0$). All terms
proportional to $\lambda^1$ are summarized by the tadpole energy
density $\varepsilon_{\rm tad}$ including the actual tadpole term as
well as the corresponding tadpole mass counterterm. The
contributions from the sunset diagram $(\propto \lambda^2)$ --
given by the correlation integral $I_1^{<}$ as well as by 
the sunset mass counterterm -- are represented by the sunset 
energy density $\varepsilon_{\rm sun}$,
\footnote{Speaking of powers of the coupling constant we mean 
the 'superficial' order of the corresponding diagrams. 
Since our self-energies $\Sigma$ are built up 
self-consistently by full Green's functions $G$ always 
higher orders of the coupling constant are resummed.} 
\bea
\varepsilon_{\rm tot}(t) & = &
\varepsilon_{\rm kin}(t) \; + \;
\varepsilon_{\rm tad}(t) \; + \;
\varepsilon_{\rm sun}(t) \; ,
\\[0.5cm]
\varepsilon_{\rm kin}(t) & = &
\phantom{-}
\frac{1}{2} \, \int \! \frac{d^{d\!}p}{(2\pi)^d} \; \;
( \, \vec{p}^{\,2} + m^2 \, ) \; \;  i \, G^{<}_{\phi \phi}(\vec{p},t,t)
\; + \;
\frac{1}{2} \, \int \! \frac{d^{d\!}p}{(2\pi)^d} \; \;
i \, G^{<}_{\pi \pi}(\vec{p},t,t) \; ,
\nnl[0.3cm]
\varepsilon_{\rm tad}(t) & = &
\phantom{-}
\frac{1}{4} \, \int \! \frac{d^{d\!}p}{(2\pi)^d} \; \;
\bar{\Sigma}_{\rm tad}(t) \; \; i \, G^{<}_{\phi \phi}(\vec{p},t,t)
\: + \:
\frac{1}{2} \, \int \! \frac{d^{d\!}p}{(2\pi)^d} \; \;
\delta\!m_{\rm tad}^2 \; \; i \, G^{<}_{\phi \phi}(\vec{p},t,t) \; ,
\nnl[0.3cm]
\varepsilon_{\rm sun}(t) & = &
 -
\frac{1}{4} \, \int \! \frac{d^{d\!}p}{(2\pi)^d} \; \; i \,
I_1^{<}(\vec{p},t,t) \: + \: \frac{1}{2} \, \int \!
\frac{d^{d\!}p}{(2\pi)^d} \; \; \delta\!m_{\rm sun}^2 \; \; i \, G^{<}_{\phi
\phi}(\vec{p},t,t) \; . 
\nn 
\eea
The calculation in Fig. \ref{plot_energy01} has been performed for
the initial distribution DT (which corresponds to a free gas of
Bose particles at temperature $T_0 = 1.59 \, m$) with a coupling
constant $\lambda/m = 16$. This state is a stationary solution
for the well-known Boltzmann equation, but not for the
Kadanoff-Baym equation. In the latter treatment the system evolves
from the uncorrelated state and the correlation energy density
$\varepsilon_{\rm sun}$ decreases with time. The decrease of the
correlation energy $\varepsilon_{\rm sun}$, which is -- with exception
of the sunset mass counterterm contribution -- initially zero, is
mainly compensated by an increase of the kinetic energy density
$\varepsilon_{\rm kin}$.
The remaining difference is equal to the change of the tadpole
energy density  $\varepsilon_{\rm tad}$ such that the total energy
density is conserved.
While the sunset and the kinetic term have always the behaviour
displayed in the figure, the change of the tadpole energy density
depends on the initial configuration and can be positive or
negative. Since the self-energies are obtained within a $\Phi$
derivable scheme (\ref{effaction}), the fundamental conservation 
laws, as e.g. energy conservation, are respected to all orders in 
the coupling constant \cite{knoll}.

\section{Comparison to the Boltzmann Limit}

As mentioned above repeatedly the Kadanoff-Baym equation
represents the full quantum field theoretical dynamics on the
one-particle level. However, its numerical solution is quite
involved and it is of strong interest to investigate, in how far
approximate schemes deviate from the full calculation. Nowadays,
transport models are widely used in the description of quantum
system out of equilibrium (cf. Introduction). Most of these models
work in the 'quasiparticle' picture, where all particles obey a
fixed energy-momentum relation and the energy  is no independent
degree of freedom anymore; it is determined by the momentum and
the (effective) mass of the particle. Accordingly, these particles
are treated with their $\delta$-function spectral shape as
infinitely long living, i.e. stable objects. This assumption is
rather questionable e.g. for high-energy heavy ion reactions,
where the particles achieve a large width due to the frequent
collisions with other particles in the high density and/or high
energy regime. Furthermore, this is doubtful for particles that
are unstable even in the vacuum. The question, in how far the
quasiparticle approximation influences the dynamics in comparison
to the full Kadanoff-Baym calculation, is of general interest
\cite{pd841,koe1}.

In order to investigate this question we formulate the Boltzmann
limit in analogy to Ref. \cite{boya}. For the detailed steps and
assumptions in the actual derivation we refer the reader to Ref.
\cite{juchem}. The Boltzmann equation 
describes the time
evolution of the momentum distribution function
\bea
N(\vec{p},t) \: = \: 
\frac{\omega_{\vec{p}}}{2} \: i \,
G^{<}_{\phi \phi}(\vec{p},t,t) \: + \:
\frac{1}{2 \omega_{\vec{p}}} \: i \,
G^{<}_{\pi \pi}(\vec{p},t,t)
\; - \;
{\rm Re\:} G^{<}_{\pi \phi}(\vec{p},t,t)
\label{boltz_def_ndist}
\eea\\[-0.2cm]
by 2 $\leftrightarrow$ 2 on-shell scattering processes, i.e.
\bea
\label{boltz_eom_ndist2}
\partial_t N(\vec{p},t)
& = & \frac{\lambda^2}{2 \omega_{\vec{p}}} \!
\int \!\!\! \frac{d^{d}\!q}{(2 \pi)^d} \!
\int \!\!\! \frac{d^{d}\!r}{(2 \pi)^d} \!
\int \!\!\! \frac{d^{d}\!s}{(2 \pi)^d} \:
\frac{1}{8 \omega_{\vec{q}} \omega_{\vec{r}} \omega_{\vec{s}}}
\: (2 \pi)^d \, \delta^{(d)}(\vec{p}\!+\!\vec{q}\!-\!\vec{r}\!-\!\vec{s}) \:
\\[0.2cm]
& \: \: & \!\!\!
\left\{
\bar{N}_{\vec{p}} \, \bar{N}_{\vec{q}} \, N_{\vec{r}} \, N_{\vec{s}}
\: - \:
N_{\vec{p}} \, N_{\vec{q}} \, \bar{N}_{\vec{r}} \, \bar{N}_{\vec{s}}
\right\}
\: \pi \:
\delta(\omega_{\vec{p}}\!+\!\omega_{\vec{q}}
\!-\!\omega_{\vec{r}}\!-\!\omega_{\vec{s}})
\phantom{aaaaaaaaaa}
\nn
\eea\\[-0.2cm]
using $N_{\vec{p}} = N(\vec{p},t)$ 
and $\bar{N}_{\vec{p}} = N(\vec{p},t) + 1$ 
for the corresponding Bose factors.

The numerical procedure for the solution of (\ref{boltz_eom_ndist2})
is basically the same as the one developed
for the solution of the Kadanoff-Baym equation.  Moreover, we
calculate the actual momentum dependent on-shell energy $\omega_{\vec{p}}$ 
for every momentum mode by a solution of
the dispersion relation including contributions from the tadpole
and the real part of the (retarded) sunset self-energy. Thus it is
guaranteed, that  the particles are treated as quasiparticles
with the correct energy-momentum relation at every time.

For the comparison between the full Kadanoff-Baym dynamics and the
Boltzmann approximation we concentrate on equilibration times. As
a measure for the equilibration time we consider the time scale on
which the initially non-isotropic distribution proceeds to the
polar symmetric equilibrium value. We define a  moment $Q(t)$
for a given momentum distribution $n(\vec{p},t)$ at time $t$ by
\bea
Q(t) \: = \:
\frac{\int \!\frac{d^{2}p}{(2 \pi)^2} \; ( p_x^2 - p_y^2 ) \; n(\vec{p},t)}
     {\int \!\frac{d^{2}p}{(2 \pi)^2}                      \; n(\vec{p},t)} \;
     ,
\eea
which vanishes for symmetric systems, e.g. for the equilibrium
state. For the Kadanoff-Baym case we calculate the actual
distribution function by
\bea n(\vec{p},t) \: = \: \sqrt{ G^{<}_{\phi \phi}(\vec{p},t,t) \;
G^{<}_{\pi \pi}(\vec{p},t,t)} \: - \: \frac{1}{2} \: . \eea
The moment $Q(t)$ shows an approximately exponential decrease in
time such that we can define a relaxation rate $\Gamma_Q$ via the
relation $ Q(T) \propto e^{- \Gamma_Q t}$.
\begin{figure}[t]
\vspace{-0.7cm}
\begin{center}
\includegraphics[width=1.0\textwidth]{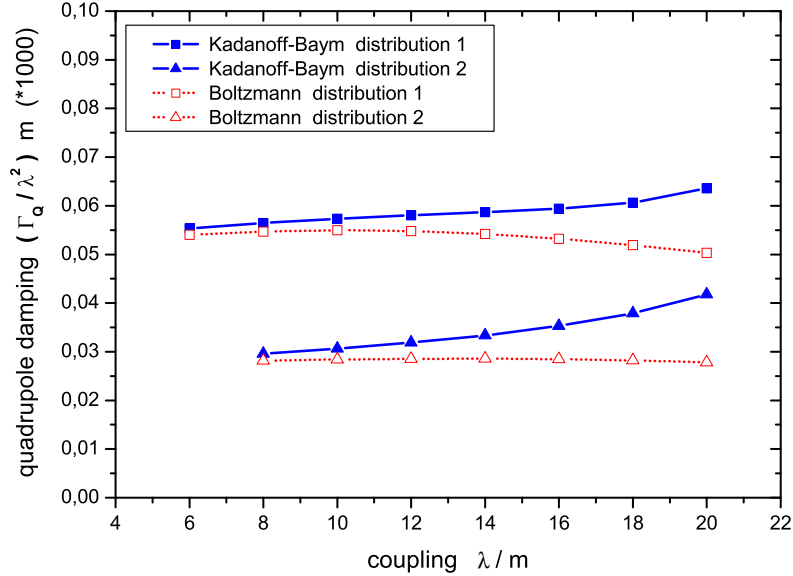}
\end{center}
\vspace{-0.7cm}
\caption[]{\it  Relaxation rate (divided by the coupling squared) for
Kadanoff-Baym and Boltzmann calculations as a function of the
interaction strength. For the two different initial configurations
the full Kadanoff-Baym evolution leads to a faster equilibration
\label{plot_quad}}
\end{figure}
In Fig. \ref{plot_quad} the relaxation rate $\Gamma_Q$ (scaled by
the coupling strength squared) is displayed for the Kadanoff-Baym
and the Boltzmann calculation for two different initial
configurations. 
\begin{figure}[hbt]
\begin{center}
\hspace*{-0.3cm}
\includegraphics[width=1.05\textwidth]{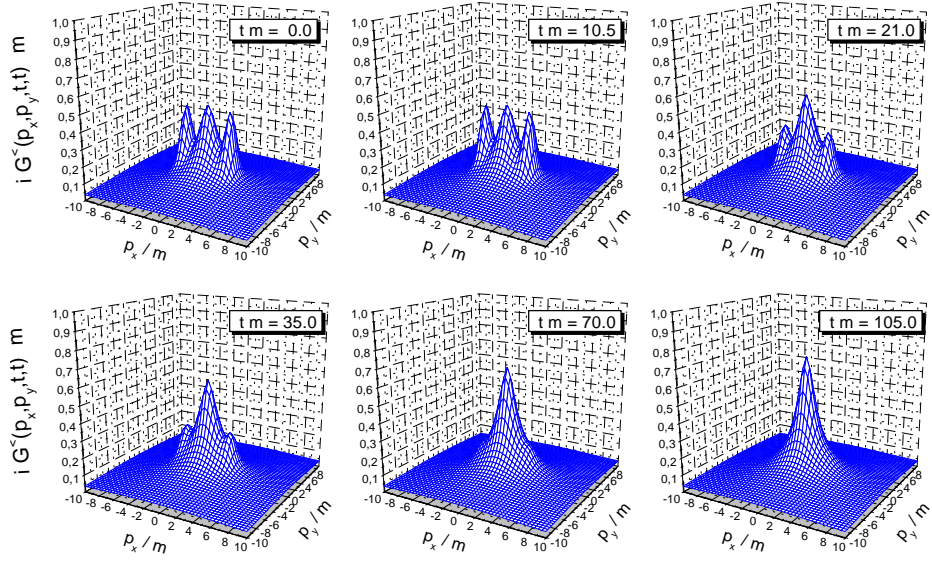}
\end{center}
\vspace*{-0.3cm}
\caption[]{\it Evolution of the Green's function in momentum space. The
equal time Green's function is displayed for various times 
${\it t \cdot m} =$ 0, 10.5, 21, 35, 70, 105. 
Starting from an initially non-isotropic shape it develops 
towards a symmetric final distribution 
\label{plot_3d}}
\end{figure}
The initial distribution 2 corresponds to the initial state
of two (on the $p_x$-axis) separated particle accummulations
(cf. Fig. \ref{plot_3d}).
The peak at low momenta again comes from the vacuum contribution
of the Green's function.
The time evolution within the full Kadanoff-Baym theory
is presented in Fig. \ref{plot_3d} by several snapshots at times
$t \cdot m =$ 0, 10.5, 21, 35, 70, 105
showing the propagation towards an isotropic (static) final state.
For distribution 1 the position and the width of the two bumps 
is modified. 

Fig. \ref{plot_quad} shows for both initializations that the 
relaxation in the full quantum calculation happens faster for 
large coupling constants than in the quasi-classical approximation, 
whereas for small couplings the equilibration times of the full and 
the approximate evolutions are comparable. 
While the scaled relaxation rate $\Gamma_Q/\lambda^2$ is nearly 
constant in the Boltzmann case, it increases with the coupling 
strength in the Kadanoff-Baym calculation (especially for 
initial distribution 2).  

Since the free Green's function -- as used in the Boltzmann calculation
-- has only support on the mass shell, only $(2 \leftrightarrow 2)$
scattering processes are described in the Boltzmann limit.
All other processes with a different number of incoming and
outgoing particles vanish. 
Within the full Kadanoff-Baym calculation this is very much 
different since here the spectral function -- determined from the 
self-consistent Green's function -- aquires a finite width. 
The Green's function has support at all energies -- although it
drops far off the mass shell.
Especially for large coupling constants, where the spectral 
function is sufficiently broad, the three particle production 
process gives a significant contribution to the collision integral. 
Since the width of the spectral function increases with the
interaction strength, such processes become more important in the 
high coupling regime. 
As a consequence the difference between both approaches is larger 
for stronger interactions as observed in Fig. \ref{plot_quad}. 
For small couplings $\lambda / m$ in both approaches basically the usual 
$2 \leftrightarrow 2$ scattering contributes and the results 
for the thermalization rate $\Gamma_Q$ are quite similar.

In summarizing this Section we point out that the full solution of
the Kadanoff-Baym equations does include $0 \leftrightarrow 4$, 
$1 \leftrightarrow 3$ and $2 \leftrightarrow 2$ off-shell collision
processes which -- in comparison to the Bolzmann on-shell
$2 \leftrightarrow 2$ collision limit -- become important when the
spectral width of the particles reaches about 1/3 of the particle
mass. 
On the other hand, the simple Boltzmann limit works
surprisingly well for smaller couplings and those cases, where the
spectral function is sufficiently narrow.

\section{Derivation of Semiclassical Transport Equations for Particles
with Dynamical Life Times}
The Kadanoff-Baym equations (5) presently cannot be solved for
general inhomogeneous problems due to the high complexity of these
equations and the huge 'computer storage' required. It is thus of
interest to obtain a semiclassical limit which is much easier to
solve e.g. by current testparticle methods.

We start again with the Kadanoff-Baym equation (5), however,
change to the Wigner representation via Fourier transformation 
of the rapidly oscillating relative coordinate $(x-y)$. 
The theory is then formulated in terms of the coordinates 
$X = (x+y)/2$ and the momentum $p$,
\bea
F_{Xp} \: = \: \int d^{\,4}(x\!-\!y) \; \; e^{i p_{\mu}
(x^{\mu}-y^{\mu})} \; \; F_{xy}.
\label{wigner_transformation}
\eea
Since convolution integrals convert under Wigner transformations
as
\bea
\int d^{\,4}(x\!-\!y) \; \; e^{i p_{\mu} (x^{\mu}-y^{\mu})} \;
\; F_{1,xz} \odot F_{2,zy}
\; = \;
e^{-i \diamond} \; F_{1,Xp} \; F_{2,Xp},
\label{wigner_convolution}
\eea
one has to deal with an infinite series in the differential
operator $\diamond$ which is a four-dimensional generalization of
the Poisson-bracket,
\bea
\diamond \, \{ \, F_{1} \, \} \, \{ \, F_{2} \, \}
\; = \; \frac{1}{2}
\left(
\frac{\partial F_{1}}{\partial X_{\mu}} \:
\frac{\partial F_{2}}{\partial p^{\mu}} \; - \;
\frac{\partial F_{1}}{\partial p_{\mu}} \:
\frac{\partial F_{2}}{\partial X^{\mu}}
\right) \; .
\label{poissonoperator}
\eea
As a standard approximation of kinetic theory only contributions
up to first order in the gradients gradient expansion 
are considered. This is
justified if the gradients in the mean spatial coordinate $X$
and momentum $p$ are small. \\
Applying this approximation scheme (Wigner transformation and
neglecting all gradient terms of order $n \ge 2$) to the
Dyson-Schwinger equations of the retarded and advanced Green's
functions one ends up with
\bea
( \, p^2 - M_{0}^{2} - {\rm Re}\Sigma^{\rm ret}_{Xp} \, ) \:
{\rm Re\,}G^{\rm ret}_{Xp}
\: & = & \:
1 \: - \: \frac{1}{4} \, \Gamma_{Xp} \:
A_{Xp} \; , \nnl[0.2cm]
( \, p^2 - M_{0}^{2} - {\rm Re}\Sigma^{\rm ret}_{Xp} \, ) \:
A_{Xp}
\: & = & \:
\Gamma_{Xp} \: {\rm Re\,}G^{\rm ret}_{Xp} \: ,
\label{dsret_firstorder}
\eea
where we have separated the retarded and advanced Green's functions
as well as the self-energies into real and imaginary contributions
\bea
G_{Xp}^{\rm ret,adv} \; = \;
{\rm Re\,}G^{\rm ret}_{Xp} \; \mp \; \frac{i}{2} \, A_{Xp} \; , \qquad
\Sigma_{Xp}^{\rm ret,adv} \; = \;
{\rm Re} \Sigma^{\rm ret}_{Xp} \; \mp \; \frac{i}{2} \, \Gamma_{Xp} \: .
\label{ret_sep}
\eea
The imaginary part of the retarded propagator is given (up to a
factor) by the normalized spectral function $A_{Xp}$ (\ref{spec_def})
while the (negative) imaginary part of the self-energy is half 
the width $\Gamma_{Xp}$. 
Above equations (\ref{dsret_firstorder}) are given for
a (renormalized) vacuum mass $M_0$ and a (renormalized)
retarded self-energy $\Sigma^{\rm ret}$. In the presence of 
an additional space-time local self-energy (e.g. tadpole self-energy)
the mass term is shifted accordingly 
$M^2_0 \rightarrow M^2_0 + \Sigma_{\rm tad}$.
From the algebraic equations (\ref{dsret_firstorder}) 
we obtain a direct relation between the real and the 
imaginary part of the propagator (provided $\Gamma_{Xp} \neq 0$):
\bea
{\rm Re\,}G^{\rm ret}_{Xp} \; = \;
\frac{p^{2} - M_{0}^{2} - {\rm Re} \Sigma^{\rm ret}_{Xp}}{\Gamma_{Xp}}
\; A_{Xp} \, .
\label{dispersion}
\eea
The algebraic solution for the spectral function shows a Lorentzian
shape with space-time and four-momentum dependent width
$\Gamma_{Xp}$. This result is valid for bosons to first order
in the gradient expansion,
\bea
A_{Xp} \; = \;
\frac{\Gamma_{Xp}} {( \,p^2 - M_{0}^{2} - {\rm Re} \Sigma^{\rm ret}_{Xp})^{2}
\: + \:
\Gamma_{Xp}^{2}/4} \; .
\label{alg_spectral}
\eea
For the real part of the retarded Green's function we get also algebraically
\bea {\rm Re\,}G^{\rm ret}_{Xp} \; = \;
\frac{p^{2} - M_{0}^{2} - {\rm Re} \Sigma^{\rm ret}_{Xp}}
 {( \,p^{2} - M_{0}^{2} - {\rm Re} \Sigma^{\rm ret}_{Xp})^{2} \: + \:
\Gamma_{Xp}^{2}/4} \, .
\label{alg_realpart}
\eea

\subsection{Transport Equations}
The Kadanoff-Baym equation (\ref{kb_spatial}) gives in the same
semiclassical approximation scheme a 
generalized transport equation,
\bea
\diamond \, \{ \, p^{2} &-& M_{0}^{2} -{\rm Re}\Sigma^{\rm ret}_{Xp} \,\} \;
            \{ \, G^{<}_{Xp} \, \} \; - \;
\diamond \, \{ \, \Sigma^{<}_{Xp} \, \} \;
            \{ {\rm Re\,}G^{\rm ret}_{Xp} \, \}
\nnl[0.0cm]
&=& \frac{i}{2} \: \left[ \:
\Sigma^{>}_{Xp} \: G^{<}_{Xp} \; - \; \Sigma^{<}_{Xp} \: G^{>}_{Xp}
\: \right] \, ,
\label{general_transport}
\eea
and a generalized mass-shell equation,
\bea
[ \, p^{2} &-& M_{0}^{2} -{\rm Re}\Sigma^{\rm ret}_{Xp} \, ] \;
G^{<}_{Xp} \; - \; \Sigma^{<}_{Xp} \; {\rm Re\,}G^{\rm ret}_{Xp}
\nnl[0.0cm]
&=&
\frac{1}{2} \,
\diamond \, \{ \, \Sigma^{<}_{Xp} \, \} \; \{ \, A_{Xp} \, \} \; - \;
\frac{1}{2} \,
\diamond \, \{ \, \Gamma_{Xp} \, \} \; \{ \, G^{<}_{Xp} \, \} \, .
\label{general_massshell}
\eea
In the transport equation (\ref{general_transport}) one recognizes
on the l.h.s. the drift term
$p^{\mu} \partial_{\mu} \bullet$ as generated by the contribution 
$\diamond\,\{ p^2 - M_0^2 \} \{\bullet \}$,
as well as the Vlasov term determined by the real
part of the retarded self-energy. On the other hand the r.h.s.
represents the collision term with its 'gain and loss' structure.
To evaluate the $\diamond \{ \Sigma^< \} \{ {\rm Re\,}G^{\rm ret} \}$-term in
(\ref{general_transport}), which does not contribute in the
quasiparticle limit, it is useful to introduce distribution
functions for the Green's functions and self-energies as
\bea
\label{sep}
i \; G_{Xp}^{<} \; = \; N_{Xp} \: A_{Xp} \: ,
\qquad \qquad
i \; G_{Xp}^{>} \; = \; (\, 1\,+\,N_{Xp} \,) \: A_{Xp} \: ,
\nnl[0.2cm]
i \; \Sigma_{Xp}^{<} \; = \; N_{Xp}^{\Sigma} \: \Gamma_{Xp} \: ,
\qquad \qquad
i \; \Sigma_{Xp}^{>} \; = \; (\, 1\,+\,N_{Xp}^{\Sigma} \,) \:
\Gamma_{Xp} \, .
\eea
Following the argumentation of Botermans and Malfliet \cite{Bot}
the distribution functions $N$ and $N^{\Sigma}$ in (\ref{sep})
should be equal in the second term of the l.h.s. of
(\ref{general_transport}) within a consistent first order
gradient expansion.
As a consequence the self-energy $\Sigma^<$ can be replaced by 
$G^< \cdot \Gamma / A$ in the term 
$\diamond \{\Sigma ^{<}\} \{{\rm Re\,}G^{\rm ret}\}$. 
The general transport equation (\ref{general_transport}) then 
can be written as
\bea
&&
\left[ \,
\diamond \, \{\, p^2 - M_0^2 -{\rm Re}\Sigma^{\rm ret}_{Xp} \,\}
            \{\, G^<_{Xp} \,\}
\, - \,
\frac{1}{\Gamma_{Xp}}
\diamond \{\Gamma_{Xp} \}
         \{(\, p^2 - M_0^2 -{\rm Re}\Sigma^{\rm ret}_{Xp} \,) \, G^<_{Xp} \,\}
\right]
\nnl[0.2cm]
&& \; \times \; \;
A_{Xp} \, \Gamma_{Xp}
\; = \; i \,
\left[ \,
\Sigma^>_{Xp} \: G^<_{Xp} \: - \: \Sigma^<_{Xp} \: G^>_{Xp} \,
\right].
\label{trans_approx}
\eea

\subsection{Test Particle Representation}

In order to obtain an approximate solution to the transport
equation (\ref{trans_approx}) we use a test particle ansatz
for the Green's function $G^{<}$, more specifically for the real and
positive semidefinite quantity
\bea
F_{Xp} = A_{Xp} \, N_{Xp} = \,
i \, G^{<}_{Xp} \: \sim \:
\sum_{i=1}^{N} \:
\delta^{(3)}\!({\vec{X}}\!\!-\!\!{\vec{X}}_i(t)) \:
\delta^{(3)}\!({\vec{p}}\!-\!{\vec{p}}_i(t)) \:
\delta(p_0\!-\!\epsilon_i(t)) . \;\;\;\;
\label{testparticle}
\eea
In the most general case (where the self-energies depend
on four-momentum $p$, time $t$ and the spatial coordinates
$\vec{X}$) the equations of motion for the test particles read
\bea
\label{eomr}
\frac{d \vec{X}_{\!i}}{dt}
\! & = & \!
\frac{1}{1 - C_{(i)}} \,
\frac{1}{2 \epsilon_i}
\left[ \,
2 \vec{p}_i + \vec{\nabla}_{\!p_i}{\rm Re}\Sigma^{\rm ret}_{(i)} +
\frac{\epsilon_i^2\!-\!\vec{p}_i^2\!-\!M_0^2\!-\!{\rm Re} \Sigma^{\rm ret}_{(i)}}
     {\Gamma_{(i)}} \: \vec{\nabla}_{\!p_i} \Gamma_{(i)}
\, \right]\!\!, \;\;\;\; \\[0.2cm]
\label{eomp}
\frac{d \vec{p}_{i}}{dt}
\! & = & \!
\frac{-1}{1-C_{(i)}} \,
\frac{1}{2 \epsilon_i}
\left[ \,
\vec{\nabla}_{\!X_i}{\rm Re}\Sigma^{\rm ret}_i +
\frac{\epsilon_i^2\!-\!\vec{p}_i^2\!-\!M_0^{2}\!-\!{\rm Re} \Sigma^{\rm ret}_{(i)}}
     {\Gamma_{(i)}} \: \vec{\nabla}_{\!X_i} \Gamma_{(i)}
\, \right]\!\!, \\[0.2cm]
\label{eome}
\frac{d \epsilon_i}{dt} \!
\! & = & \!
\frac{1}{1 - C_{(i)}} \,
\frac{1}{2 \epsilon_i}
\left[ \,
\frac{\partial{\rm Re}\Sigma^{\rm ret}_{(i)}}{\partial t} +
\frac{\epsilon_i^2\!-\!\vec{p}_i^2\!-\!M_0^{2}\!-\!{\rm Re} \Sigma^{\rm ret}_{(i)}}
     {\Gamma_{(i)}} \,
\frac{\partial \Gamma_{(i)}}{\partial t}
\, \right]\!\!,
\eea
where the notation $F_{(i)}$ implies that the function is taken at
the coordinates of the test particle at time $t$, i.e.
$F_{(i)} \equiv F(t,\vec{X}_{i}(t),\vec{p}_{i}(t),\epsilon_{i}(t))$.

In (\ref{eomr}-\ref{eome}) a common multiplication factor
$(1-C_{(i)})^{-1}$ appears, which contains the energy derivatives
of the retarded self-energy
\bea \label{correc} C_{(i)} \: = \: \frac{1}{2 \epsilon_i} \left[
\frac{\partial}{\partial \epsilon_i} \,{\rm Re}\Sigma^{\rm ret}_{(i)} \: +
\: \frac{\epsilon_i^2 - {\vec p}_i^2 - M_0^2 - {\rm Re}
\Sigma^{\rm ret}_{(i)}}{\Gamma_{(i)}} \: \frac{\partial }{\partial
\epsilon_i} \, \Gamma_{(i)} \right] \: . \eea
It yields a shift of the system time $t$ to the 'eigentime' of
particle $i$ defined by $\tilde{t}_{i} = t /(1-C_{(i)})$. As the
reader immediately verifies, the derivatives with respect to the
'eigentime', i.e. $d \vec{X}_i / d \tilde{t}_i$, $d \vec{p}_i / d
\tilde{t}_i$ and $d \epsilon_i / d \tilde{t}_i$ then emerge
without this renormalization factor for each test particle $i$ when
neglecting higher order time derivatives in line with the
semiclassical approximation scheme. In the limiting case
of particles with vanishing gradients of the width $\Gamma_{Xp}$
these equations of motion  reduce to the well-known transport
equations of the quasiparticle picture.

Following Refs. \cite{caju} we take $M^{2} = p^2 - {\rm Re\,}\Sigma^{\rm ret}$
as an independent variable instead of $p_0$, which
then fixes the energy (for given $\vec{p}$ and $M^{2}$) to
\bea p_{0}^{2} \; = \; \vec{p}^{2} \: + \: M^{2} \: + \:
{\rm Re\,}\Sigma_{X\vec{p}M^2}^{\rm ret} \, .
\label{energyfix}
\eea
Eq. (\ref{eome}) then turns to
\bea
\label{eomm}
\frac{d (M_i^2-M_0^2)}{dt} \; = \;
\frac{M_i^2 - M_0^2}{\Gamma_{(i)}} \;
\frac{d \Gamma_{(i)}}{dt}
\eea
for the time evolution of the test particle $i$ in the invariant
mass squared as derived in Refs. \cite{caju}.
For applications of the semiclassical off-shell transport
approach we refer the reader to Refs. \cite{caju}.

\section{Summary}
In this article we have studied the time evolution of an
interacting field theoretical system, i.e.  $\phi^4$-field theory
in 2+1 space-time dimensions,  on the basis of the Kadanoff-Baym
equations for a spatially homogeneous system including the
self-consistent tadpole and sunset self-energies. We find that
equilibration is achieved only by inclusion of the sunset 
self-energy. Simultaneously, the time evolution of the single-particle
spectral function has been calculated for various initial conditions.
A comparison of the full solution of the Kadanoff-Baym equations
with the solution for the corresponding Boltzmann equation shows
that a consistent inclusion of the spectral function has a sizeable
impact on the equilibration rates if the width of the spectral
function becomes larger than $\sim$1/3 of the particle mass.

Furthermore, the conventional transport
of particles in the on-shell quasiparticle limit has been extended to
particles of finite life time by means of a dynamical spectral
function $A(X,\vec{p},M^2)$. Starting again from the Kadanoff-Baym
equations we have derived in consistent first order gradient expansion
equations of motion for test particles with respect to their time
evolution in $\vec{X}, \vec{p}$ and $M^2$. This off-shell
propagation has been examined for a couple of model cases in Refs.
\cite{caju} as well as for nucleus-nucleus collisions showing
that -- at subthreshold energies -- the off-shell dynamics play an 
important role for the production of energetic particles.

\vspace{0.3cm} \noindent
{\bf Acknowledgements} \\
\noindent
The authors acknowledge valuable discussions with C. Greiner and
S. Leupold throughout these studies.


\begin{thebibliography}{8.}
\addcontentsline{toc}{section}{References}

\bibitem{js61}
   J. Schwinger, 
   J. Math. Phys. \textbf{2}, 407 (1961).
%
\bibitem{Wang1}
   S. J. Wang and W. Cassing, 
   Ann. Phys. (N.Y.) \textbf{159}, 328 (1985).
%
\bibitem{CaWa}
   W. Cassing and S. J. Wang, 
   Z. Phys. A \textbf{337}, 1 (1990).
%
\bibitem{cs85}
   K. Chou, Z. Su, B. Hao, and L. Yu, 
   Phys. Rept. \textbf{118}, 1 (1985).
%
\bibitem{md90}
   S. Mr\'{o}wczy\'{n}ski and P. Danielewicz, 
   Nucl. Phys. B \textbf{342}, 345 (1990).
%
\bibitem{mh94}
   S. Mr\'{o}wczy\'{n}ski and U. Heinz, 
   Ann. Phys. (N.Y.) \textbf{229}, 1 (1994).
%
\bibitem{Stoecker}
   H. St\"ocker and W. Greiner, 
   Phys. Rept. \textbf{137}, 277 (1986).
%
\bibitem{Bertsch}
   G. F. Bertsch and S. Das Gupta, 
   Phys. Rept. \textbf{160}, 189 (1988).
%
\bibitem{CMMN}
   W. Cassing, V. Metag, U. Mosel, and K. Niita,
   Phys. Rept. \textbf{188}, 363 (1990).
%
\bibitem{Cass}
   W. Cassing and U. Mosel, 
   Prog. Part. Nucl. Phys.  \textbf{25}, 235 (1990).
%
\bibitem{Fuchs}
   C. Fuchs and T. Gaitanos,
   nucl-th/0211091; 
   this volume.
%
\bibitem{URQMD}
   S. Bass, M. Belkacem, M. Bleicher et al.,
   Prog. Part. Nucl. Phys. \textbf{41}, 255 (1998).
%
\bibitem{CB99}
   W. Cassing and E. L. Bratkovskaya, 
   Phys. Rept. \textbf{308}, 65 (1999).
%
\bibitem{kb62} L. P. Kadanoff and G. Baym,
   {\it Quantum statistical mechanics}, Benjamin, New York, 1962.
%
\bibitem{pd841}
   P. Danielewicz, 
   Ann. Phys. (N.Y.) \textbf{152}, 239 (1984); {\it ibid.} 305.
%
\bibitem{Bot}
   W. Botermans and R. Malfliet, 
   Phys. Rept. \textbf{198}, 115 (1990).
%
\bibitem{Mal}
   R. Malfliet, 
   Prog. Part. Nucl. Phys. \textbf{21}, 207 (1988).
%
\bibitem{ph95}
   P. A. Henning, 
   Nucl. Phys. A \textbf{582}, 633 (1995); 
   Phys. Rept. \textbf{253}, 235 (1995);
   this volume.
%
\bibitem{gl98}
   C. Greiner and S. Leupold, 
   Ann. Phys. (N.Y.) \textbf{270}, 328 (1998).
%
\bibitem{Zuo}
   S. J. Wang, W. Zuo, and W. Cassing, 
   Nucl. Phys. A \textbf{573}, 245 (1994).
%
\bibitem{CNW}
   W. Cassing, K. Niita, and S. J. Wang, 
   Z. Phys. A \textbf{331}, 439 (1988).
%
\bibitem{Rudy1}
   R. Malfliet, 
   Nucl. Phys. A \textbf{545}, 3 (1992).
%
\bibitem{Rudy2}
   R. Malfliet, 
   Phys. Rev. B \textbf{57}, R11027 (1998).
%
\bibitem{Pavel}
   P. Danielewicz and S. Pratt, 
   Phys. Rev. C \textbf{53}, 249 (1996).
%
\bibitem{Mora1}
   V. \v{S}pi\v{c}ka, P. Lipavsk\'y, and K. Morawetz, 
   Phys. Rev. B \textbf{55}, 5095 (1997); 
   Phys. Lett. A \textbf{240}, 160 (1998);
   this volume.
%
\bibitem{Mora2}
   P. Lipavsk\'y, V. \v{S}pi\v{c}ka, and K. Morawetz, 
   Phys. Rev. E \textbf{59}, 1291 (1999).
%
\bibitem{Mora3}
   P. Lipavsk\'y, K. Morawetz, and V. \v{S}pi\v{c}ka, 
   Annales de Physique \textbf{26}, 1 (2001).
%
\bibitem{Peter} 
   A. Peter et al., 
   Z. Phys. A \textbf{358}, 91 (1997);
   Z. Phys. C \textbf{71}, 515 (1997).
%
\bibitem{berges1}
    J. Berges and J. Cox,
    Phys. Lett. B \textbf{517}, 369 (2001).
%
\bibitem{berges2}
    J. Berges and G. Aarts,
    Phys. Rev. D \textbf{64}, 105010 (2001).
%
\bibitem{berges3}
    J. Berges,
    Nucl. Phys. A \textbf{699}, 847 (2002).
%
\bibitem{knoll}
    Y. B. Ivanov, J. Knoll, and D. N. Voskresensky,
    Nucl. Phys. A \textbf{657}, 413 (1999).
%
\bibitem{knollren}
    H. van Hees and J. Knoll,
    Phys. Rev. D \textbf{65}, 025010 (2002);
    Phys. Rev. D \textbf{65}, 105005 (2002).
%
\bibitem{koe1}
    H. S. K\"ohler,
    Phys. Rev. C \textbf{51}, 3232 (1995).
%
\bibitem{boya}
    D. Boyanovsky, I. D. Lawrie, and D.-S. Lee,
    Phys. Rev. D \textbf{54}, 4013 (1996).
%
\bibitem{juchem} S. Juchem, W. Cassing, and C. Greiner, 
    hep-ph/0307353.
%
\bibitem{caju}
    W. Cassing and S. Juchem,
    Nucl. Phys. A \textbf{665}, 377 (2000);
    Nucl. Phys. A \textbf{672}, 417 (2000);
    Nucl. Phys. A \textbf{677}, 445 (2000).


\end{thebibliography}
\end{document}